\documentstyle[12pt]{article}
\begin{document}
\title{ Possible manifestations of the graviton
 background }
\author
{By Michael A. Ivanov \\
Chair of Physics, \\
Belarus State University of Informatics and Radioelectronics, \\
6 P. Brovka Street, BY 220027, Minsk, Belarus.  \\
E-mail: ivanovma@gw.bsuir.unibel.by}
\date{June 6, 2002}
\maketitle
\begin{abstract}
Possible effects are considered which would be caused by a hypothetical
superstrong interaction of photons or massive bodies with single gravitons
of the graviton background. If full cosmological redshift magnitudes are
caused by the interaction, then the luminosity distance in a flat
non-expanding universe as a function of redshift is very similar to
the specific function which fits supernova cosmology data
by Riess et al. From another side, in this case every massive body,
slowly moving relatively to the background, would experience a constant
acceleration, proportional to the Hubble constant, of the same order
as a small additional acceleration of Pioneer 10, 11. \\
\end{abstract}


\section[1]{Introduction }

In the standard cosmological model \cite {1}, as well as in inflationary
cosmological models \cite{29}, redshifts of remote objects
are explained by expansion of the universe. A model of expansion
gives an exact dependence of a distance $r$ from an
observer to a source on a redshift $z.$ There is a known uncertainty of
estimates of the Hubble constant $H$ because of difficulties to establish
a scale of cosmological distances which is independent on redshifts \cite
{2,3}.  Today, as one could think,
there are not obvious observant facts which would demand some
alternative model to interpret an origin of redshifts. But one cannot
exclude that the effect may have some non-dopplerian nature.
\par
In alternative cosmological models, which are known as "tired-light" ones,
the cosmological redshift is considered namely as a non-dopplerian effect.
Several mechanisms for photon energy loss have been supposed \cite {11,12}.
There exist different opinions, what a cosmological model makes the better fit
to the existing astrophysical data on some kinds of cosmological tests
(compare, for example, \cite {13,14} with \cite {12}).
\par
  In this paper, possible manifestations of the graviton
background in a case of hypothetical superstrong gravitational quantum
interaction are considered. From one side, the author brings the reasons
that a quantum interaction
of photons with the graviton background would lead to redshifts of remote
objects too. The author considers a hypothesis about  an existence of the
graviton background to be independent from the standard
cosmological model. One cannot affirm that such an interaction is the
only cause of redshifts. It is possible, that the one gives a small
contribution to an effect magnitude only.  But we cannot exclude that such
an interaction with the graviton background would be enough to explain the
effect without an attraction of the big bang hypothesis. Comparing the own
model predictions with supernova cosmology data by Riess et al \cite{12a},
the author finds here good accordance between the redshift model and
observations.
\par
From another side, it is shown here, that every massive body, with a non-zero
velocity $v$ relatively to the isotropic graviton background,
should experience
a constant acceleration. If one assumes that a full observable redshift
magnitude is caused by such a quantum interaction with single gravitons,
then this acceleration will have
the same order of magnitude as a small additional acceleration of NASA
deep-space probes (Pioneer 10/11, Galileo, and Ulysses), about which it was
reported by Anderson's team \cite{15}.
\par
It is known, that a gravitational interaction between two particles
is very weak on big distances.
One may expect, that its non-dimensional coupling
"constant", which could be an analogue of QED's coupling
constant $\alpha \simeq
1/137,$  would be proportional to $E_{1}E_{2}/E_{Pl}^{2},$ where $E_{1}$
and $E_{2}$ are energies of particles, $E_{Pl} \simeq 10^{19} GeV$ is the
Planck energy (i.e. the mentioned "constant" is a bilinear function of
energies of
particles). May such an interaction with gravitons decelerate a big cosmic
probe or, at the worst, give observable redshifts? We must take into
account, that we know little of quantum gravity (see, for example,
\cite{17}). Today, there does not exist a complete theory of it.
The weak field limit
is successfully investigated in  the context of linearized gravity \cite{19}.
In this approach, one considers gravitons without self-interaction, comparing
their energies to the Planck scale. Unified theories, including gravity,
contain, as a rule, big spectra of non-observed particles \cite{31,32}.
\par
The Newton gravitational constant $G$ characterizes an interaction on a macro
level. But on this level, from a quantum point of view, the interaction may be
superstrong. For example, if we consider two stars, having the Sun masses,
as "particles", then, for this case, the non-dimensional "constant" will be
equal to $10^{72}.$ Of course, it means only, that one cannot consider an
interaction between such "particles" as a result of exchange by single
gravitons. Because of self-interaction of gravitons, possible Feynman's
diagrams should be complex and should contain a lot of crossing chains of
vertexes. Because of it, the Newton constant $G$ may be, perhaps, much
smaller than an unknown constant which characterizes a single act of
interaction.
\par
All considered effects depend on the equivalent temperature $T$  of
the graviton background, which are unknown out of standard cosmological
models, based on the big bang hypothesis. But we must take into account,
that known estimates of a classical
gravitational wave background intensity are consistent with
values of this equivalent temperature, which may be
not more than few Kelvin degrees \cite{20,21,30}. Probably,
future gravitational
wave detectors (for the low frequencies $\sim 10^{-3} Hz$)
will give more exact estimates \cite{22} - \cite{25}.

\section[2]{ Photon energy losses  due to an interaction with the graviton
background }

Let us introduce the hypothesis, which is considered here to be independent
from the
standard cosmological model: there exists the isotropic graviton background.
Then  photon scattering is possible on gravitons
$\gamma + h \to \gamma + h,$ where $\gamma $ is a photon and $h$ is a
graviton, if one of the gravitons is virtual. The energy-momentum
conservation law prohibits energy transfer to free gravitons.
\par
 Average energy losses of a photon with an energy  $E $ on a way
$dr $ will be equal to
\begin{equation}
                  dE=-aE dr,
\end{equation}
where $a$ is a constant.
Here we take into account
that a gravitational "charge" of a photon must be proportional
to $E $ (it gives the factor  $E^{2}$ in a cross-section) and a
normalization of a photon wave function gives the factor $E^{-1}$ in
the cross-section. Also we assume that a photon average energy loss
$\bar \omega $ in one act of interaction is relatively small to a
photon energy $E. $ We must identify $a=H/c,$ where $c$ is the light
velocity, to have the Hubble law for small distances \cite{4}.
\par
A photon energy $E$ should depend on a distance from a source $r$ as
\begin{equation}
                      E(r)=E_{0} \exp(-ar),
\end{equation}
where $E_{0}$ is an initial value of energy.
\par
The expression (2) is just only so far as the condition $\bar \omega
<< E(r)$ takes place. Photons with a very small energy may loss or acquire
an energy changing their direction of propagation after scattering.  Early
or late such photons should turn out in thermodynamic equilibrium with the
graviton background, flowing into their own background. Decay of virtual
gravitons should give photon pairs for this background too. Possibly,
we know the last one as the cosmic microwave background \cite {33,34}.
\par
It follows from the expression (2) that an exact dependence $r(z)$ is the
following one:
\begin{equation}
                     r(z)= ln (1+z)/a,
\end{equation}
if an interaction with the graviton background is the only cause of redshifts.
We see that this redshift do not depend
on a light frequency. For small $z,$ the dependence $r(z)$ will be linear.
\par
The expressions (1) - (3) are the same that appear in other tired-light
models (compare with \cite {12}). In our case, they follow from
a possible existence of the isotropic graviton background,
from quantum electrodynamics, and from the fact that a gravitational "charge"
of a photon must be proportional to $E.$
\par

\section[3]{An additional relaxation of a photon flux due to non-forehead
collisions with gravitons}

\par
An interaction of photons with the graviton background will lead to an
additional relaxation of a photon flux, caused by transmission of a momentum
transversal component to some photons. Photon flux's average energy losses
on a way $dr$ should be proportional to $badr,$ where $b$ is a constant of the
order $1.$ These losses are connected with a rejection of a part of photons
from a source-observer direction. Such the relaxation together with the
redshift will give a connection between visible object's diameter and its
luminosity (i.e. the ratio of an  object visible
angular diameter to a square root of visible luminosity),
distinguishing from the one of the standard cosmological model.
\par
Let us consider that in a case of a non-forehead collision of a graviton
with a photon, the latter leaves a photon flux detected
by a remote observer (an assumption of narrow
beam of rays). Then we get the following estimate for the factor $b:$
\begin{equation}
                                b=3/2 + 2/\pi = 2,137.
\end{equation}
It is assumed here that a cross-section of  interaction is modified by the
 factor
$\vert \cos \alpha \vert$ where $\alpha$ is an angle between wave vectors of a
photon and of a graviton raiding on it from front or back hemispheres.
To average on the angle $\alpha,$ one must take into account a dependence
of a graviton flux, which falls on a picked out area (cross-section), on
the angle $\alpha.$
Thus in the simplest case of the uniform non-expanding universe with the
Euclidean space, we shall have the quantity
$$(1+z)^{(1+b)/2} \equiv (1+z)^{1,57}$$
in a visible object diameter-luminosity connection if  whole redshifts
would caused by such an interaction with the background (instead of
$(1+z)^{2}$ for the expanding uniform universe). Of course, this quantity
may be modified with evolutionary effects. For near
sources, the estimate of the factor $b$ will be an increased one.
\par

\section[4a]{Comparison of the redshift model with supernova cosmology data}

In a case of flat no-expanding universe, a photon flux relaxation can be
characterized by the factor $b,$ so that the luminosity distance $D_{L}$
\cite{12a} is equal in our model to:
\begin{equation}
D_{L}=a^{-1} \ln(1+z)\cdot (1+z)^{(1+b)/2} \equiv a^{-1}f_{1}(z;b),
\end{equation}
where $z$ is a redshift. The theoretical estimate for $b$ is: $b= 3/2+2/\pi
=2.137.$ Thus, the redshift
\begin{equation}
z=\exp(ar)-1
\end{equation}
and the luminosity distance
$D_{L}$
are characterized in the model by two parameters: $H$ and $b$ ($r$ is a
geometrical distance). One can introduce an effective Hubble constant
\begin{equation}
H_{eff} \equiv c{dz}/{dr}.
\end{equation}
In our model
\begin{equation}
H_{eff} = H \cdot (z+1);
\end{equation}
in a language of expansion it can be interpreted as "a current deceleration
of the expansion".
\par
High-z Supernova Search Team data \cite{12a} give us a possibility
to evaluate $H$
in our model. Instead of prompt fitting to data, we can use one of the best
fits of the function $D_{L}(z; H_{0},\Omega_
{M},\Omega_{\Lambda})$ to supernovae data
from \cite{12a} (see Eq.2 in \cite{12a}) with $\Omega_{M}=-0.5$ and
$\Omega_{\Lambda}=0,$ which is unphysical in the original work. For $1-
\Omega_{M}>0$ and $1+\Omega_{M}z>0,$ the function $D_{L}(z; H_{0},\Omega_
{M},\Omega_{\Lambda})$ is equal to (see the integral in \cite{12b}):
$$D_{L}=a^{-1}(1+z) m^{-1} \sinh (\ln \vert {(k-m)}/{(k+m)} \vert-\ln
\vert {(1-m)}/{(1+m)}\vert)\equiv$$
\begin{equation}
\equiv a^{-1}f_{2}(z; \Omega_{M},\Omega_{\Lambda}),
\end{equation}
 where $m \equiv (1-\Omega_{M})^{1/2}, k \equiv (1+\Omega_{M}z)
^{1/2}.$
Assuming $b=2.137,$ we can find $H$ from the connection:
\begin{equation}
HD_{L}/H_{0}D_{L}=f_{1}(z;b)/f_{2}(z; \Omega_{M},\Omega_{\Lambda}),
\end{equation}
where $H_{0}$ is an estimate of the Hubble constant from \cite{12a}
(see Table 1).
We see that $H/H_{0} \simeq const,$ a deviation $(H-<H>)/<H>$
from an average value $<H>
\simeq 1.09H_{0}$ is less than $\pm 5\%.$

\begin{table}[t]
\caption{Comparison with supernovae data}
\begin{tabular}{cccccccccccc} \hline
$z$      &0  &0.1  &0.2  &0.3  &0.4  &0.5  &0.6  &0.7  &0.8  &0.9  &1.0\\ \hline
$f_{1}$&0&0.110&0.242&0.396&0.570&0.765&0.983&1.222&1.480&1.759&2.058\\ \hline
$f_{2}$&0&0.103&0.219&0,359&0.511&0.677&0.863&1.074&1.301&1.565&1.854\\ \hline
$H/H_{0}$&-  &1.068&1.105&1.103&1.115&1.130&1.139&1.138&1.138&1.124&1.110\\
\hline
\end{tabular}
\end{table}

It means, that the model is in good accordance with supernovae data. This
accordance cannot become worse, if one evaluates both of the parameters fitting
our two-parametric function $D_{L}(z;H,b)$ to supernovae data.
\par
If one would suggest that $f_{1}(z;b)$ describes results of observations in
an expanding universe, one could conclude that it is "an accelerating one". But
a true conclusion may be strange: our universe is not expanding, and redshifts
have the non-dopplerian nature.

\par
\section[4]{ Other possibilities to verify the conjecture about redshift's local
nature}

If redshifts of remote objects would be provided by such the local cause as
an interaction of photons with the graviton background, then a spectrum of
ultrastable laser radiation after a delay line should have a small redshift
too. It gives us a hope to carry out a straight verification of the
conjecture about redshift's local nature on the Earth after creation of
ultrastable lasers with an instability $\sim 10^{-17}$ \cite{26} and of
optical delay lines for a delay $\sim 10$ s \cite{5}.
\par
A discrete character of photon energy losses by interaction with
gravitons may produce a specific deformation of a spectrum of
ultrastable laser radiation in a delay line: a step would appear beside
a spectral line, from the side of low frequencies \cite{8}.
Such steps would appear beside single narrow spectral lines of remote
objects too.  A width of the step should linear raise with $z$.
For remote objects, this additional effect would be caused by multifold
interactions of a small part of photons with the graviton background.
This effect would be observable, if $\bar \omega$ will be comparable
with a spectral line width, a redshift of which one will measure.
\par
An establishment of a cosmological distance scale, which
will be independent
of redshifts, would allow to verify the expression (3) or its consequence:
\begin{equation}
                                r_{1}/ r_{2}= \ln (1+z_{1})/ \ln (1+z_{2}),
\end{equation}
where $r_{1}$ and $ r_{2}$ are the distances to the sources 1 and 2,
$ z_{1} $ and $ z_{2}$ are their redshifts.
\par
It follows from (6) for small $ar$ that
\begin{equation}
                      z = ar + (ar)^{2}/2 + (ar)^{3}/6 + \dots,
\end{equation}
where $a = H/c.$ Estimates of coefficients by $r^{2}, r^{3}, \dots,
$ which would be received from an analysis of astrophysical data for small
$z,$ could be compared with their values from (12) (see \cite{9}). The
Canada-France redshift survey \cite{10} may serve as an example of big
statistics which could make possible such a comparison.
\par
We can verify a proportionality of the ratio of an  object visible
angular diameter to a square root of visible luminosity to the quantity
$(1+z)^{1,57},$
which takes place in the assumption that the uniform no-expanding universe
with the
quasi-Euclidean space are realized. We must keep in the mind, that
evolutionary effects would change a value of the ratio.
\par
Perspective programs of big statistics accumulation for quasar redshifts
on a base of the microlensing effect \cite{6} would be useful to verify the
local nature of redshifts, too.
\par

\section[5]{Deceleration of massive bodies by the graviton background }

It was reported by Anderson's team \cite{15} , that NASA deep-space
 probes (Pioneer 10/11, Galileo, and Ulysses) experience a small additional
constant acceleration, directed towards the Sun. Today, a possible origin of
the effect is unknown. It must be noted here that the reported
direction of additional
acceleration may be a result of the simplest conjecture, which was accepted
by the authors to provide a good fit for all probes. One should compare
different conjectures to choose the one giving the best fit.
\par
We consider here a deceleration of massive bodies, which would give a similar
deformation of cosmic probes' trajectories. The one would be
a result of interaction of a massive body with the graviton background, but
such an additional acceleration will be directed against a body velocity.
\par
It follows from an universality of gravitational interaction, that not only
photons, but all other objects, moving relatively to the background, should
loss their energy too due to such a quantum interaction with gravitons. If
$a=H/c,$ it turns out that massive bodies must feel a constant  deceleration
of the same order of magnitude as a small additional acceleration of
cosmic probes.
\par
Let us now denote as $E$ a full energy of a moving body which has a velocity
$v$ relatively to the background. Then energy losses of the body by an
interaction with the graviton background (due to forehead collisions with
gravitons)
on the way $dr$ must be expressed by the same formula (1):
$$ dE=-aE dr,$$
where $a=H/c.$ If $dr=vdt,$ where $t$ is a time, and
$E=mc^{2}/\sqrt{1-v^{2}/c^{2}},$ we get for the body acceleration
$w \equiv dv/dt$ by a non-zero velocity:
\begin{equation}
w = - ac^{2}(1-v^{2}/c^{2}).
\end{equation}
We assume here, that non-forehead collisions with gravitons give only
stochastic deviations of a  massive body's velocity direction, which are
negligible.
For small velocities:
\begin{equation}
w \simeq - Hc.
\end{equation}
If the Hubble constant $H$ is equal to $1.6 \cdot 10^{-18} s^{-1},$
the acceleration will be equal to
\begin{equation}
w \simeq - 4.8 \cdot 10^{-10} m/s^{2},
\end{equation}
that corresponds approximately to one half of the observed additional
acceleration for NASA probes.
\par
We must emphasize here that the acceleration $w$ is directed against a body
velocity only in a special system of reference (in which the graviton background
is isotropic). In other systems of reference, we will find its direction,
using transformation formulae for an acceleration (see \cite{9}).
We can assume that the graviton background and the microwave one are
isotropic in one system of reference (the Earth velocity relatively to
the microwave background was determined in \cite{16}).
\par
To verify our conjecture about an origin of probes' additional acceleration,
one could re-analyze radio Doppler data for probes. One should find a
velocity of the special system of reference and a constant probe acceleration
$w$ in this
system which must be negative, as it is described above. These two parameters
must provide the best fit for all probes, if our conjecture is true.  In such
a case, one can get an independent estimate of the Hubble constant, based on
the measured value of probe's additional acceleration: $H= \mid w \mid
/c.$
\par
Under influence of such a small additional acceleration $w$, a probe must
move on a deformed trajectory.  Its view will be determined by small seeming
deviations from exact conservation laws for energy and angular momentum
of a not-fully reserved body system which one has in a case of neglecting
with the graviton background. For example, Ulysses should go some nearer to
the Sun when the one rounds it. It may be interpreted as an additional
acceleration, directed towards the Sun, if we shall think that one deals
with a reserved body system.
\par
It is very important to understand, why such an acceleration has not been
observed for planets. This acceleration will have different directions by
motion of a body on a closed orbit.  As a result, an orbit should be deformed.
Possibly, the general relativity effect of a perihelion revolution \cite{28}
would lead to a partial compensation of an average influence of the considered
acceleration within a big time. This question needs a further consideration.

\section[6]{Estimates of a  cross-section and of new constants which would
characterize an interaction with single gravitons  }

Let us assume that a full redshift magnitude is caused by an interaction with
single gravitons. If $\sigma (E,\omega)$ is a cross-section of interaction by
forehead collisions of
a photon with an energy $E$ with a graviton, having an energy $\omega,$
we consider factually (see (1)), that
$${d \sigma (E,\omega) \over E d \Omega} = const(E),$$
where $d \Omega$ is a space angle element, and the function $const(x)$
has a constant value for any $x$.
If $f(\omega,T)d \Omega /2 \pi$ is a spectral density of graviton flux in
the limits of $d \Omega$ in some direction,
i.e. an intensity of a graviton flux is equal to an integral
$ (d \Omega/2 \pi) \int_{0}^{\infty}
f(\omega,T)d \omega,$ $T$ is an equivalent temperature of the graviton
background, we can write for the Hubble constant $H=ac,$
introduced in the expression (1):
$$H={1 \over 2\pi} \int_{0}^{\infty} \frac {\sigma (E,\omega)}{E}
f(\omega,T)d \omega.$$
If $f(\omega,T)$ can be described by the Planck formula for equilibrium
radiation, then $$ \int_{0}^{\infty} f(\omega,T)d \omega
= \sigma T^{4},$$ where $\sigma$ is the Stephan-
Boltzmann constant \cite{27}.
As carriers of a gravitational "charge" (without consideration of
spin properties), gravitons should be described in the same manner as
photons (compare with \cite{19}), i.e. one can write for them:
$${d \sigma (E,\omega) \over \omega d\Omega} = const(\omega).$$
Now let us introduce a new dimensional constant $D$ so, that for
forehead collisions:
$$\sigma (E,\omega)= D \cdot E \cdot \omega. $$
Then
\begin{equation}
H= {1 \over 2\pi} D \cdot \bar \omega \cdot (\sigma T^{4}),
\end{equation}
where $\bar \omega$ is an average graviton energy. \footnote{In 
this version, the remainder of this section is replaced with a 
corrected fragment}
\par
Assuming $T
\sim 3 K, \bar \omega \sim 10^{-4} eV,$ and $H = 1.6 \cdot
10^{-18} s^{-1},$ we get the following estimate for $D:$ $$D \sim
10^{-27} m^{2}/eV^{2},$$ that gives us the phenomenological
estimate of cross-section by the same $E$ and $\bar \omega$:
$$\sigma (E,\bar \omega) \sim 10^{-35} m^{2}.$$ One can compare
this value with the cross-section of quasi-elastic
neutrino-electron scattering \cite{18}, having, for example, the
order $\sim 10^{-44} m^{2}$ by a neutrino energy about $6\ GeV.$
\par
Let us introduce new constants: $G_{0}, l_{0}, E_{0},$ which are
analogues, on this new scale, of classical constants: the Newton
constant $G,$ the Planck length $l_{Pl},$ and the Planck energy
$E_{Pl}$ correspondingly. Let $$D \equiv (l_{0}/ E_{0})^{2} =
(G_{0}/c^{4})^{2},$$ where $ l_{0}=\sqrt{ G_{0} \hbar /c^{3}},
E_{0}=\sqrt{ \hbar c^{5}/ G_{0}}.$ Then we have for these new
constants: $$G_{0} \sim 1.6 \cdot 10^{39} m^{3}/kg \cdot s^{2},
l_{0} \sim 2.4\cdot 10^{-12} m, E_{0} \sim 1.6\ KeV.$$ If one would
replace $G$ with $G_{0},$ then an electrostatic force, acting
between two protons, will be  $\sim 2\cdot 10^{13}$ times smaller
than a gravitational one by the same distance.
\par
Using $E_{0}$ instead of $E_{Pl},$ we can evaluate the new
non-dimensional "constant" (a bilinear function of $E$ and
$\omega$) $k,$ which would characterize one act of interaction: $k
\equiv E \cdot \omega / E_{0}^{2}. $ We must remember here, that
an universality of gravitational interaction allows to expect that
this floating coupling "constant" $k$ should characterize
interactions of any particles with an energy $E,$ including
gravitons, with single gravitons. For $E \sim 1 eV$ and $\omega
\sim 10^{-4} eV,$ we have $k  \sim 4 \cdot 10^{-9}.$ But for  $E
\sim 25 MeV$ and  $\omega \sim 10^{-3} eV,$ we shall have $k  \sim
10^{-2},$ i.e. $k$  will be comparable with QED's constant
$\alpha.$ Already by $E \sim \omega \sim 5 KeV,$ such an
interaction would have the same intensity as a strong interaction
($k \sim 10$).

\section[7]{Conclusion }

Independently from the described conjecture, we would wait that a straight
verification of redshift's nature on the Earth should be one of main works for
coming ultrastable lasers. In a case of the dopplerian nature of redshifts,
one will get a negative result for a laser beam frequency shift after a delay
line. Such a negative result would be an important indirect experimental
confirmation of the big bang hypothesis. Today for most people, a positive
result seems to be impossible. But in a case of such an unexpected positive
result, the redshift laser experiment would be a key one for cosmology.
\par
One can wait that unification of gravity with physics of particles will
need non-ordinary solutions, for example, introduction of many-dimensional
spaces, in which a model of gravity has the basic symmetries of the Standard
Model \cite{35}. From another side, the author feels a necessity to include
gravity in the model of composite fermions to describe a set of generations
and to solve a problem of particle masses \cite{36}.
\par
If further investigations display that an anomalous NASA probes'
acceleration cannot be explained by some technical causes, left out of
account today, it will give a big push to a further development of
physics of particles. Both supernova cosmology data and the Anderson's team
discovery may change a gravity position in a hierarchy of known interactions,
and, possibly, give us a new chance to unify their description.

\par
This paper earlier version's one-page abstract was poster presented at the
Particles and Nuclei International Conference (PANIC'99), June 10-16, 1999,
Uppsala, Sweden.
\par

\end{document}